\documentclass[twoside,11pt,letter]{article}
\usepackage[margin=1.2in]{geometry}
\usepackage{physics}
\usepackage{times}
\usepackage{amsmath}
\usepackage{amsfonts}
\usepackage{amssymb}
\usepackage{amsthm}
\usepackage{makeidx} 
\usepackage{graphicx}
\usepackage{mathrsfs}
\usepackage{tabularx}
\usepackage{caption}
\usepackage{bbold}
\usepackage{color}
\usepackage{fancybox}
\usepackage{verbatim}
\usepackage{hyperref}
\usepackage{tikz}
\usepackage{float}
\usepackage{url}
\usetikzlibrary{arrows,automata}
\usetikzlibrary{trees}
\usetikzlibrary{shadows}
\usepackage[framemethod=TikZ]{mdframed}
\usepackage{mdframed}
\usepackage[ruled,vlined]{algorithm2e}
\usepackage{booktabs}
\usepackage{youngtab}
\mdfdefinestyle{theorem}{%
linecolor=black,
outerlinewidth=.5pt,
roundcorner=5pt,
innertopmargin=.5\baselineskip,
innerbottommargin=.75\baselineskip,
innerrightmargin=20pt,
innerleftmargin=20pt,
backgroundcolor=green!20,
shadow=true,
shadowcolor=green!20}
\mdfdefinestyle{algo}{%
linecolor=black,
outerlinewidth=.5pt,
roundcorner=5pt,
innertopmargin=.5\baselineskip,
innerbottommargin=.75\baselineskip,
innerrightmargin=10pt,
innerleftmargin=10pt,
backgroundcolor=yellow!20,
shadow=true,
shadowcolor=yellow!20}
\mdfdefinestyle{example}{%
linecolor=black,
outerlinewidth=.5pt,
leftline = false,
rightline = false,
innertopmargin=.5\baselineskip,
innerbottommargin=.75\baselineskip,
innerrightmargin=10pt,
innerleftmargin=10pt,
backgroundcolor=white}

\def\b1{\mathbb{1}}

\def\e1{\mathsf{1}}

\def\eE{\mathsf{E}}

\def\eP{\mathsf{P}}

\def\sF{{\mathscr{F}}}

\def\sX{{\mathscr{X}}}

\def\cov{\mathsf{Cov}}

\def\var{\mathsf{Var}}

\def\e1{\mathsf{1}}

\def\of0{(0)}

\def\bf0{\mathbf{0}}
\def\cp1{\mathbb{CP}^1}

\newtheorem{theorem}{Theorem}[section]

\theoremstyle{definition}
\newtheorem{definition}[theorem]{Definition}
\theoremstyle{definition}

\newtheorem{proposition}[theorem]{Proposition}

\begin{document}

\title{\textbf{Mortgage Burnout and Selection Effects\\
in Heterogeneous Cox Hazard Models}}
\author{\textbf{Andrew Lesniewski}\\
Department of Mathematics\\
Baruch College\\
One Bernard Baruch Way\\
New York, NY 10010\\
USA}

\maketitle

\begin{abstract}
We study the aggregate hazard rate of a heterogeneous population whose individual event intensities are modeled as Cox (doubly stochastic) processes. In the deterministic hazard setting, the observed pool hazard is the survival weighted mean of the individual hazards, and its time derivative equals the mean individual hazard drift minus a variance term. This yields a transparent structural explanation of burnout in mortgage pools. We extend this perspective to stochastic intensity models. The observed pool hazard remains a survival-weighted mean, but now evolves as an Ito process whose drift contains the mean drift of the individual hazards and a negative selection term driven by cross-sectional dispersion, together with a diffusion term inherited from the common factor. We formulate the general identity and discuss special cases relevant to mortgage prepayment modeling.
\end{abstract}

\newpage

\section{Introduction}

A standard feature of mortgage prepayment modeling is the \emph{burnout effect}: conditional on a favorable refinancing environment, the observed prepayment rate of a mortgage pool tends to decline with seasoning because the most refinance sensitive borrowers exit first. This phenomenon is widely documented in the mortgage-backed securities literature \cite{DM81}, \cite{ST89}, \cite{DL14}. Empirically, burnout is one of the most robust features of mortgage prepayment data. The standard explanation is borrower heterogeneity. Borrowers differ in transaction costs, financial sophistication, mobility, credit constraints, and other factors affecting their propensity to refinance. When interest rates decline, borrowers with the strongest refinancing incentives tend to prepay first, leaving behind a population with progressively lower propensity to refinance. As a result, the observed pool prepayment rate declines over time even if the aggregate refinancing incentive remains unchanged.

In heterogeneous deterministic hazard models this mechanism admits a simple structural representation,
\begin{equation*}
\frac{d}{dt}\,\bar{\lambda}(t)=\eE_t(\dot{\lambda}(t,x))-\var_t(\lambda(t,x)),
\end{equation*}
where $\bar{\lambda}(t)$ is the observed pool hazard and $\eE_t$, $\var_t$ are computed under the survival weighted cross-sectional distribution of borrower types.

The identity above has an interesting conceptual interpretation. The evolution of the pool hazard can be viewed as a selection effect in a heterogeneous population. Borrowers with higher hazard rates tend to exit the pool earlier, so that the surviving population becomes progressively enriched in lower-hazard types.  

This mechanism is mathematically analogous to the \emph{Price equation} from evolutionary biology \cite{P70}, which decomposes the change in the mean value of a trait into a selection component and a transmission component. In its simplest form, the Price equation states that the change in the population mean of a trait $z_i$ satisfies
\begin{equation*}
\Delta \bar z=\frac{\cov(w_i,z_i)}{\bar w}+\frac{1}{\bar w}\, \eE(w_i \Delta z_i),
\end{equation*}
where $w_i$ denotes fitness. 

In the present context, the role of the trait is played by the borrower hazard $\lambda(x)$, while the role of fitness is played by survival. The negative variance term in the burnout identity is precisely the continuous-time analog of the covariance selection term in the Price equation. Thus mortgage burnout can be interpreted as a form of natural selection within a heterogeneous population of borrowers: high-hazard borrowers prepay first, leaving behind a population with progressively lower average hazard. Importantly, the burnout effect therefore arises generically in heterogeneous hazard models and does not require behavioral assumptions about borrower refinancing decisions.

The purpose of this note is to formulate a stochastic analog of this identity when each borrower-level hazard is itself a Cox process.

\section{Mortgage Pool Heterogeneity}

Let $(\Omega,\sF,(\sF_t)_{t\ge0},\eP)$ be a filtered probability space describing the time evolution of the pool of mortgages, and let $(X,\sX,F)$ be a measurable type space representing the unobserved borrower heterogeneity. For each $x\in X$, let $\lambda_t(x)$ be a nonnegative $\sF_t$-adapted intensity process. Typically, $\lambda_t$ depends explicitly on a number of loan characteristics $y$, such as refinance incentive (the difference between the coupon on the loan and the current mortgage rate), size, age, loan to value ratio, FICO score, geographic region, which represent the observed heterogeneity. Unless stated otherwise, we will suppress this dependence for the sake of notational convenience.

Conditionally on the filtration, the survival process of a borrower of type $x$ is
\begin{equation*}
S_t(x)=\exp\Big(-\int_0^t \lambda_s(x)\,ds\Big).
\end{equation*}
The aggregate survival process of the pool is
\begin{equation*}
\bar{S}_t=\int_X S_t(x)\,F(dx).
\end{equation*}

\begin{definition}
\textit{The observed pool hazard $\bar{\lambda}_t$ is defined by
\begin{equation*}
\bar{S}_t=\exp\Big(-\int_0^t \bar{\lambda}_s\,ds\Big).
\end{equation*}
Equivalently,
\begin{equation*}
\bar{\lambda}_t=-\frac{d}{dt}\,\log\bar{S}_t .
\end{equation*}}
\end{definition}
Since
\begin{equation*}
dS_t(x)=-\lambda_t(x)S_t(x)\,dt,
\end{equation*}
it follows that
\begin{equation*}
\bar{\lambda}_t=\frac{\int_X \lambda_t(x)S_t(x)\,F(dx)}{\int_X S_t(x)\,F(dx)}\,.
\end{equation*}

\begin{definition}
\textit{Define the survival weighted cross-sectional measure
\begin{equation*}
F_t(dx)=\frac{S_t(x)F(dx)}{\int_X S_t(y)\,F(dy)} .
\end{equation*}
For any integrable function $\phi(x)$,
\begin{equation*}
\eE_t(\phi)=\int_X \phi(x)\,F_t(dx),
\end{equation*}
is the expected value with respect to $F_t$ and $\var_t(\phi)$ is the variance with respect to $F_t$.}
\end{definition}

Taking the time derivative of $\log \bar{S}_t$, we find that
\begin{equation}
\bar{\lambda}_t=\eE_t(\lambda_t(x)).
\end{equation}

The survival weighted measure $F_t$ evolves over time because borrowers with higher hazard rates exit the pool earlier. This induces a selection effect in expectations taken under $F_t$.

Let $\phi_t(x)$ be a semimartingale indexed by borrower type $x$. A direct differentiation of
\begin{equation*}
\eE_t(\phi_t)=\frac{\int_X \phi_t(x)S_t(x)\,F(dx)}{\int_X S_t(x)\,F(dx)}
\end{equation*}
yields the identity
\begin{equation}
d\eE_t(\phi_t)=\eE_t(d\phi_t)-\cov_t(\lambda_t(x),\phi_t(x))\,dt .
\end{equation}

This identity may be viewed as a continuous-time analog of the \emph{Price equation} from evolutionary biology, which decomposes the change in the population mean of a trait into a direct evolution component and a selection component. In the present setting, borrower survival plays the role of fitness, and the covariance term captures the selective removal of high-hazard borrowers from the pool.

\section{Deterministic Hazards}

If each borrower hazard $\lambda(t,x)$ is differentiable in time, differentiation yields the following \emph{deterministic burnout identity}.

\begin{theorem}
Suppose each borrower hazard $\lambda(t,x)$ is differentiable in time. Then the pool hazard satisfies
\begin{equation}
\frac{d}{dt}\,\bar{\lambda}(t)=\eE_t(\dot{\lambda}(t,x))-\var_t(\lambda(t,x)).
\end{equation}
\end{theorem}

The second term represents a selection effect: borrowers with higher hazard rates exit the pool earlier, reducing the average hazard of the surviving population. In particular, if individual hazards are constant in time, $\dot{\lambda}(t,x)=0$, then
\begin{equation}
\begin{split}
\frac{d}{dt}\,\bar{\lambda}(t)&=-\var_t(\lambda(t,x))\\
&\leq 0.
\end{split}
\end{equation}
Thus heterogeneity alone produces a monotone decline in the observed pool hazard.

\section{Stochastic Hazards}

\subsection{Common Factor Hazard Dynamics}

Suppose each borrower hazard follows an Ito process
\begin{equation}
d\lambda_t(x)=\mu_t(x)\,dt+\sigma_t(x)\,dW_t.
\end{equation}
Idiosyncratic borrower shocks diversify in large pools and therefore do not contribute to the aggregate hazard dynamics. Since the pool hazard satisfies
\begin{equation*}
\bar{\lambda}_t=\eE_t(\lambda_t(x)),
\end{equation*}
the selection identity yields the following stochastic generalization of the burnout formula.

\begin{proposition}\textit{Under the common factor dynamics above, the pool hazard evolves according to
\begin{equation}
d\bar{\lambda}_t=\Big(\eE_t(\mu_t(x))-\var_t(\lambda_t(x))\Big)dt+\eE_t(\sigma_t(x))\,dW_t .
\end{equation}}
\end{proposition}

The drift contains the same negative variance term 	$-\var_t(\lambda_t(x))$ as in the deterministic burnout identity. Thus heterogeneity generates a structural downward pressure on the observed pool hazard even when individual hazards follow stochastic dynamics.

Similar selection effects arise in heterogeneous credit hazard models, where dispersion in firm level default intensities leads to declining average hazard among surviving firms.

\subsection{Measure Change Interpretation}

The survival weighted distribution $F_t$ can also be interpreted as a change of measure induced by the survival process. Define the weighting factor
\begin{equation*}
H_t(x)=\frac{S_t(x)}{\bar{S}_t}\,.
\end{equation*}

Then expectations under the survival weighted distribution satisfy
\begin{equation*}
\eE_t(\phi)=\int_X \phi(x) H_t(x) F(dx).
\end{equation*}
Thus $F_t$ may be viewed as a Radon-Nikodym tilt of the original type distribution $F$ with density proportional to survival. From this perspective the burnout identity arises because the weighting factor $H_t(x)$ evolves according to
\begin{equation*}
\frac{d}{dt}\log H_t(x)=-\lambda_t(x)+\bar{\lambda}_t .
\end{equation*}
Consequently the change in expectations under the tilted measure contains a covariance term
\begin{equation*}
-\cov_t(\lambda_t(x),\phi_t(x)),
\end{equation*}
which represents the selective removal of high hazard borrowers from the population.

This representation shows that burnout can be interpreted as a measure change phenomenon: the observed pool dynamics correspond to expectations taken under a survival biased distribution of borrower types. In other words, the cross-sectional distribution of borrower types observed in a seasoned mortgage pool is endogenously tilted toward lower hazard borrowers. Burnout therefore arises as a purely statistical selection effect rather than a behavioral change in borrower refinancing incentives.

\section{Frailty Factor Models}

A commonly used mortgage specification is the \emph{frailty model} \cite{VMS79}, \cite{DJ08}. In mortgage applications the frailty factor captures unobserved borrower characteristics such as refinancing costs, mobility, financial sophistication, or credit constraints. It assumes that borrower hazards factor into a borrower specific frailty parameter and a common stochastic intensity $\lambda^0_t(y)$ depending on the observed features $y$,
\begin{equation*}
\lambda_t(x, y)=f(x)\lambda^0_t(y),
\end{equation*}
where $f(x)>0$ represents borrower heterogeneity. Then
\begin{equation*}
\bar{\lambda}_t(y)=\lambda^0_t(y) \eE_t(f(x)),
\end{equation*}
and
\begin{equation*}
\var_t(\lambda_t(x, y))=\lambda^0_t(y)^2\, \var_t(f(x)).
\end{equation*}
The burnout dynamics are therefore governed by the evolution of the survival weighted distribution of $f(x)$.

We now illustrate these ideas with several explicit examples of the distribution $F$. In mortgage applications the gamma and lognormal frailty models are particularly natural. The gamma model is analytically tractable and produces a hyperbolic burnout profile, while the lognormal model arises naturally from multiplicative borrower characteristics such as refinancing incentive, credit quality, and mobility factors.

\subsection{Gamma Frailty}

The gamma frailty model is widely used in survival analysis because it admits closed form expressions. Suppose the initial frailty distribution is
\begin{equation*}
f \sim \mathrm{Gamma}(k,\theta),
\end{equation*}
with mean $k\theta$ and variance $k\theta^2$. If $\lambda^0_t(y)=\lambda$ is constant, then
\begin{equation*}
S_t(f)=\exp(-\lambda f t).
\end{equation*}
The survival weighted distribution remains gamma,
\begin{equation*}
f \mid \text{survival at }t \sim\mathrm{Gamma}\Big(k,\,\frac{\theta}{1+\theta \lambda t}\Big).
\end{equation*}

Hence
\begin{equation*}
\eE_t(f)=\frac{k\theta}{1+\theta \lambda t}.
\end{equation*}
The observed hazard becomes
\begin{equation*}
\bar{\lambda}_t=\lambda\, \frac{k\theta}{1+\theta \lambda t}.
\end{equation*}
Thus burnout produces a hyperbolic decay of the pool hazard.

\begin{proposition}
\textit{Under gamma frailty with constant common factor $\lambda$, the heterogeneous Cox population is observationally equivalent to a deterministic hazard model with time-varying pool hazard
\begin{equation}
\bar{\lambda}_t=\frac{\bar{\lambda}_0}{1+\theta \lambda t}\,.
\end{equation}}
\end{proposition}

\subsection{Lognormal Frailty}

Suppose
\begin{equation*}
f=\exp(Y), \qquad Y\sim N(\mu,\sigma^2).
\end{equation*}
Then the survival weighted distribution satisfies
\begin{equation*}
F_t(df)\propto\exp(-\lambda f t)\,\frac{1}{f\sigma\sqrt{2\pi}}\,\exp\Big(-\frac{(\log f-\mu)^2}{2\sigma^2}\Big) df .
\end{equation*}
Closed-form expressions are not available, but the pool hazard
\begin{equation*}
\bar{\lambda}_t=\lambda \eE_t(f)
\end{equation*}
can be evaluated numerically.

For small dispersion $\sigma$, a Laplace approximation yields
\begin{equation*}
\bar{\lambda}_t\approx\lambda\exp\Big(\mu+\frac{\sigma^2}{2}\Big)\exp\Big(-\sigma^2 \lambda t\Big),
\end{equation*}
which produces approximately exponential burnout.

\subsection{Normal Frailty}

Strictly speaking, frailty variables in survival models are required to be nonnegative, since they scale the baseline hazard.  For this reason the gamma and lognormal distributions are commonly used in applications.  The normal distribution does not satisfy this positivity constraint and is therefore not
appropriate as a structural frailty model.  Nevertheless, it is useful as a simple analytic benchmark illustrating how cross-sectional dispersion generates
burnout.

For illustration we therefore also consider the normal distribution.
\begin{equation*}
f \sim N(m,s^2),
\end{equation*}
with $f>0$ implicitly assumed. Then
\begin{equation*}
\bar{\lambda}_t=\lambda\,\frac{\int f e^{-\lambda f t}\phi(f)\,df}{\int e^{-\lambda f t}\phi(f)\,df}.
\end{equation*}
For small variance $s^2$, a second-order expansion gives
\begin{equation*}
\bar{\lambda}_t \approx\lambda\big(m - s^2 \lambda t\big),
\end{equation*}
showing a linear burnout effect for short horizons.

\subsection{Multivariate Frailty}

In many mortgage applications borrower heterogeneity is driven by several unobserved factors rather than a single scalar frailty variable.  A natural extension is therefore a \emph{multivariate frailty model} \cite{DJ08}, in which the borrower hazard takes the form
\begin{equation*}
\lambda_t(x,y)=f(x)^\top \lambda^0_t(y),
\end{equation*}
where $f(x)=(f_1(x),\ldots,f_k(x))^\top$ represents a vector of borrower-specific frailty factors and $\lambda^0_t(y)$ is a vector of common factor intensities that may depend on the observed loan characteristics $y$. Examples of such latent factors include borrower mobility, refinancing transaction costs, credit constraints, and other behavioral characteristics that are difficult to observe directly but influence the propensity to prepay.

Under this specification the pool hazard becomes
\begin{equation*}
\bar{\lambda}_t(y)=\eE_t(f(x))^\top \lambda^0_t(y),
\end{equation*}
where $\eE_t(f(x))$ denotes the survival weighted mean frailty vector. The burnout dynamics are therefore governed by the evolution of the survival weighted distribution of the multivariate frailty factors. Applying the deterministic burnout identity yields
\begin{equation*}
\frac{d}{dt}\,\bar{\lambda}_t(y)=\eE_t(\dot{\lambda}_t(x,y))-\lambda^0_t(y)^\top \cov_t(f(x))\,\lambda^0_t(y),
\end{equation*}
where $\cov_t(f(x))$ is the survival weighted covariance matrix of the frailty vector.

Thus in the multivariate setting the burnout effect is driven by the cross-sectional covariance structure of borrower frailties.  Directions in factor space with larger dispersion contribute more strongly to the decline of the pool hazard as high frailty borrowers exit the pool. This formulation shows that mortgage burnout can be interpreted as a selection effect operating on a multidimensional latent factor structure
for borrower behavior.

\subsection{Connection with Cox Proportional Hazard Models}

The representations considered above can be viewed as a particular parameterization of Cox proportional hazard models. In the classical Cox framework, the hazard is written as
\begin{equation*}
\lambda_t(y) = \lambda^{\mathrm{base}}_t\,\exp\big(\beta^\top y_t\big),
\end{equation*}
where $\lambda^{\mathrm{base}}_t$ is a baseline hazard and $y_t$ denotes observable covariates.

In the present setting, the roles of baseline hazard and covariates may be interpreted differently. The frailty component $\lambda_t(x)$, or more generally $f(x)$ in the factor specification, can be viewed as a borrower-specific random modification of the baseline hazard, while the common factor $\lambda^0_t(y)$ plays the role of a time-varying systematic component driven by observable characteristics.

Under this interpretation, heterogeneous Cox intensity models correspond to Cox-type specifications with random, possibly multidimensional, baseline heterogeneity. The burnout effect then arises from the endogenous evolution of the cross-sectional distribution of these latent baseline components under survival weighting. This perspective highlights that burnout is not tied to a particular functional form of the hazard, but rather reflects a general selection mechanism operating in Cox-type models with latent heterogeneity.

\section{Conclusion}

The observed hazard of a heterogeneous Cox population remains a survival-weighted average of borrower hazards. Under common-factor stochastic dynamics, the observed pool hazard satisfies
\begin{equation*}
d\bar{\lambda}_t=\Big(\eE_t(\mu_t(x))-\var_t(\lambda_t(x))\Big)dt+\eE_t(\sigma_t(x))\,dW_t .
\end{equation*}
The negative variance term provides a structural stochastic analog of burnout: heterogeneity induces a systematic downward drift in the observed hazard through selection effects. The resulting formula may be interpreted as a stochastic continuous time analog of the Price equation, showing that burnout arises from a universal selection mechanism in heterogeneous hazard populations.

\bibliographystyle{plain}

\end{document}